\documentclass[aps,twocolumn,pra,superscriptaddress,amsmath,showpacs,tightenlines]{revtex4-1}
\usepackage{amssymb}
\usepackage{amsmath}
\usepackage{graphicx}
\usepackage{subfigure}
\usepackage{natbib}
\usepackage{epsfig}
\usepackage{amsfonts}
\usepackage{mathrsfs}
\usepackage{CJK}
\usepackage{xcolor}
\usepackage{threeparttable}
\usepackage[toc,page,title,titletoc,header]{appendix}

\begin{document}

\title{Controlling photons by phonons via giant atom in a waveguide QED setup}

\author{Xinyu Li}
\affiliation{Center for Quantum Sciences and School of Physics, Northeast Normal University, Changchun 130024, China}
\author{Wei Zhao}
\affiliation{Center for Quantum Sciences and School of Physics, Northeast Normal University, Changchun 130024, China}
\author{Zhihai Wang}
\email{wangzh761@nenu.edu.cn}
\affiliation{Center for Quantum Sciences and School of Physics, Northeast Normal University, Changchun 130024, China}

\begin{abstract}
We investigate the single photon scattering in a phonon-photon hybrid system in the waveguide QED scheme. In our consideration, an artificial giant atom, which is dressed by the phonons in a surface acoustic wave resonator, interacts with a coupled resonator waveguide (CRW) nonlocally via two connecting sites. Together with the interference effect by the nonlocal coupling, the phonon serves as a controller to the transport of the photon in the waveguide. On the one hand, the coupling strength between the giant atom and the surface acoustic wave resonator modulates the width of the transmission valley or window in the near resonant regime. On the other hand, the two reflective peaks induced by the Rabi splitting degrade into a single one when the giant atom is large detuned from the surface acoustic resonator, which implies an effective dispersive coupling. Our study paves the way for the potential application of giant atoms in the hybrid system.
\end{abstract}

\date{\today}

\maketitle
\section{Introduction}
Waveguide quantum electrodynamics (QED)~\cite{Gu2017,Roy2017} mainly studies the interaction between the limited light field in the waveguides and matter at the quantum level.  Due to the achievable strong coupling  between light and matter, the superconducting circuit provides an ideal platform for realizing and exploring the physical properties of waveguide QED~\cite{Kannan2020}, such as resonance fluorescence~\cite{OA2010,ICHOI2011,ICHOI2015}, collective Lamb shifts~\cite{PYWEN2019} and Dicke super-radience and sub-radiance~\cite{RHD1954,AFVAN2013,MM2019}. Meanwhile, as the carrier of the information, the propagation of the photon in the waveguide can be controlled by a two or three-level system. In such a manner, the single and few photon scattering has attracted lots of interest, which is aiming to design coherent quantum devices, for example,  quantum transistors~\cite{ZHOU2008}, routers~\cite{ZHOU2013} and frequency converters~\cite{WANG2014}.

In the conventional quantum optics scenario, the size of the natural atom, whose radius is in the order of $10^{-10}$\,m, is much smaller than the wavelength of the photons ($\lambda \approx 10^{-7}-10^{-6}$\,m) in the resonator or waveguide, therefore the dipole approximation is usually applied by considering that the electromagnetic field is uniform in the atomic regime~\cite{KOCKUM2020}. However, a pioneering experimental work in 2014 suggests that the superconducting qubit can be coupled to the phonon field in the surface nonlocally~\cite{MVGUS2014}. Such a system is named as ``giant atom'' by Kockum. The giant atom exhibits some interesting quantum effects which do not exist in the small atom setup, such as frequency dependent Lamb shift~\cite{PYWEN2019}, non-Markovian oscillation bound state~\cite{guo2020,SLONGHI2021,QIU2023,lim2023}, chiral physics~\cite{wang2021,SORO2022,WANG2022,LIU2022,WANGX2022} and so on. Meanwhile, the giant atom model is also proposed in the cold atom system~\cite{AGT2019} and synthetic dimension~\cite{du2022,xiao2022}. Recently, the giant atom with more than two coupling points or the coupling between more giant atoms and the waveguide has also been realized in the superconducting circuits~\cite{ANDERSSON2019,AMV2021,WANGZQ2022}.

The artificial superconducting qubit, which serves as a giant atom, can simultaneously couple to the phonon and photon in the microwave frequency. Therefore, via the data bus supplied by the giant atom, it is possible to design the photon-phonon hybrid system, to realize the mutual control between the photons and phonons. In this letter, we propose such a model in the context of the waveguide QED as shown in Fig.~\ref{f1}. Unlike Delsing's experiment~\cite{MVGUS2014} where the giant atom couples to the surface acoustic wave (SAW) nonlocally, we here limit the SAW in a resonator, which locally couples to the two-level system (named as atom in what follows). To demostrate the effect of the giant atom, we further couple the atom to a CRW, which supports the propagation of microwave photons. In this way, we show how to control the transport of the photon by tuning the degree of freedom of the phonon in the SAW resonator.

When the atom and the SAW resonator resonantly couple to each other, we demonstrate the Rabi splitting by the single photon reflective spectrum and a stronger atom-SAW resonator coupling strength is beneficial to widen the reflection valley. Due to the photonic interference between the two atom-waveguide connecting points, we also observe a transmission window under some certain conditions, and the window can also be widened by increasing the atom-SAW resonator coupling. Therefore, the giant atom supplies us with an unconventional way to manipulate the scattering of photons in the waveguide.

\section{MODEL AND HAMILTONIAN}
\label{model}

\begin{figure}
\begin{centering}
\includegraphics[width=1\columnwidth]{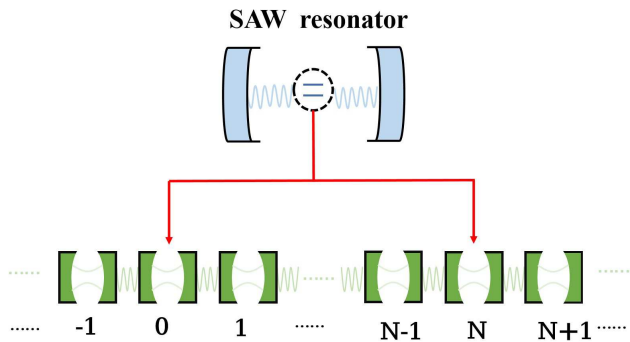}
\par\end{centering}
\caption{Schematic diagram of simultaneous coupling of a giant atom with a SAW resonator and a CRW.}
\label{f1}
\end{figure}

As schematically shown in Fig.~\ref{f1}, the system we consider is composed of a two-level system which is coupled to both a SAW resonator and a one-dimensional CRW with infinite length.
The two-level system, which serves as a giant atom,  couples to the waveguide nonlocally via two separate sites.  The Hamiltonian $\mathcal{H}$ of the system can be divided into three parts, i.e., $\mathcal{H}=\mathcal{H}_{0}+\mathcal{H}_{c}+\mathcal{H}_{I}$.
The first part is (Hereafter, we set $\hbar=1$).
\begin{equation}
\label{eq:1}
\mathcal{H}_{0}=\omega _0a^\dagger a+\Omega \left|e\right\rangle \left\langle e\right|+\lambda \left ( \sigma ^+a+a ^\dagger\sigma ^-  \right ),
\end{equation}
which describes the coupling between the giant atom and the SAW resonator. This coupling can be realized via the interdigital transducer (IDTs)~\cite{ZENG2021,DATTA1986,MDS2007,CHU2018}. Here, $a$ is the annihilation operator of the SAW resonator with frequency $\omega_0$, $\Omega$ is the transition frequency
of the giant atom between its ground state $\left|g\right\rangle $ and the excited state $\left|e\right\rangle $.
As a reference, we have set the frequency of the ground state $\left|g\right\rangle $ as $\omega_g=0$.
$\sigma ^{+}=(\sigma ^{-})^{\dagger}=\left |e  \right \rangle\left \langle g \right |$ is the raising operator.
The real number $\lambda$ is the magnitude of the coupling constant between the giant atom and the SAW resonator. In the above Hamiltonian, we have used the rotating wave approximation by considering the parameter regime of $\lambda\ll(\omega_0,\Omega)$.

The second part $\mathcal{H}_{c}$ of the Hamiltonian $\mathcal{H}$ represents the free Hamiltonian of the CRW, which can be written as
\begin{equation}
\label{eq:2}
\begin{split}
\mathcal{H}_c=\omega _c\sum_{j}^{}b_j^\dagger b_j-\xi \sum_{j=-\infty}^{+\infty}\left (b_{j+1}^\dagger b_j +b_j^\dagger b_{j+1} \right ),
\end{split}
\end{equation}
Here $\omega_c$ is the frequency of the resonators, and $b_j$ is the boson annihilation operator on site $j$.
$\xi$ is the hopping strength between the nearest neighbour resonators.

The third part $\mathcal{H}_I$ of the Hamiltonian describes the coupling between the CRW and the giant atom via the $0$th and $N$th sites.
Under the rotating wave approximation, the Hamiltonian $\mathcal{H}_I$ can be written as
\begin{equation}
\label{eq:3}
\mathcal{H}_I=g\left ( b_0^\dagger  \sigma ^ - + b_0  \sigma ^ +\right )+g\left ( b_N^\dagger  \sigma ^ - + b_N  \sigma ^ +\right ),
\end{equation}
where $g$ is the coupling strength between the CRW and the giant atom,  and has been assumed as a real number.

\section{SINGLE-PHOTON SCATTERING }
\label{scattering}

In this section, we will discuss the behavior of the single-photon scattering.
We consider that a single-photon with wave vector $k$ is incident from the left side of the CRW.
Since the excitation number in the system is conserved, the eigenstate in the single-excitation subspace can be written as
\begin{equation}
\label{eq:4}
\left |\psi  \right \rangle=\left ( v_aa^\dagger +u_e\sigma ^++\sum_{j}^{}c_j b_j^\dagger  \right )\left |G  \right \rangle,
\end{equation}
where $\left |G  \right \rangle$ is the ground state of the whole hybrid system.
The parameters $v_a$ and $u_e$ are the excitation amplitudes of the SAW resonator and the giant atom, respectively. $c_j$ is the probability amplitude for finding a photon excited in the $j$th resonator of the CRW.
In the regimes of $j<0$ and $j>N$, the amplitude $c_j$ can be written in the form
\begin{equation}
\label{eq:5}
\begin{split}
c_j=\begin{cases}e^{ikj}+re^{-ikj},&j<0
\\te^{ikj},&j>N,
\end{cases}
\end{split}
\end{equation}
where $r$ and $t$ are respectively the single-photon reflection and transmission amplitudes.
Hereafter, the wave vector $k$ is considered to be dimensionless by setting the distance between the two nearest resonators to be unity.
In the regime covered by the giant atom, the photon propagates back and forth, and the amplitude $c_j$ for $0\leqslant j\leqslant N$ can be expressed as
\begin{equation}
\label{eq:6}
c_j=Ae^{ikj}+Be^{-ikj}.
\end{equation}
Solving the Schr\"{o}dinger equation $H\left |\psi  \right \rangle=E\left |\psi  \right \rangle$ in the region of $j\neq0,N$ yields a dispersion relationship of $E=\omega _c-2\xi\cos k$.
Furthermore, the continuity conditions at $j=0$ and $j=N$ tells us $1+r=A+B$ and $Ae^{ikN}+Be^{-ikN}=te^{ikN}$, respectively.
Combining the above formulas, the reflection rate $R=\left | r \right |^2$ can be obtained as
\begin{equation}
\label{eq:7}
R=\frac{4g^4 \Delta_k^2\cos ^4 {\frac{kN}{2}}}{4g^4\Delta_k^2\cos ^2 {\frac{kN}{2}}+\xi^2Q^2\sin^2{k}+2\xi g^2Q\Delta_k\sin{k}\sin{kN}},
\end{equation}
where  $\Delta =E-\Omega$ is the detuning between the giant atom and the propagating photons in the CRW,  $\Delta_k =\omega _0-E$ is the detuning between the propagating photons in the CRW and the SAW resonator, and the function $Q$ is defined as $Q=\left [ \Delta \left ( \Delta +\Omega -\omega _0 \right ) -\lambda ^2\right ]$.

\begin{figure}[ht]
\centering
\includegraphics[width=0.49\columnwidth]{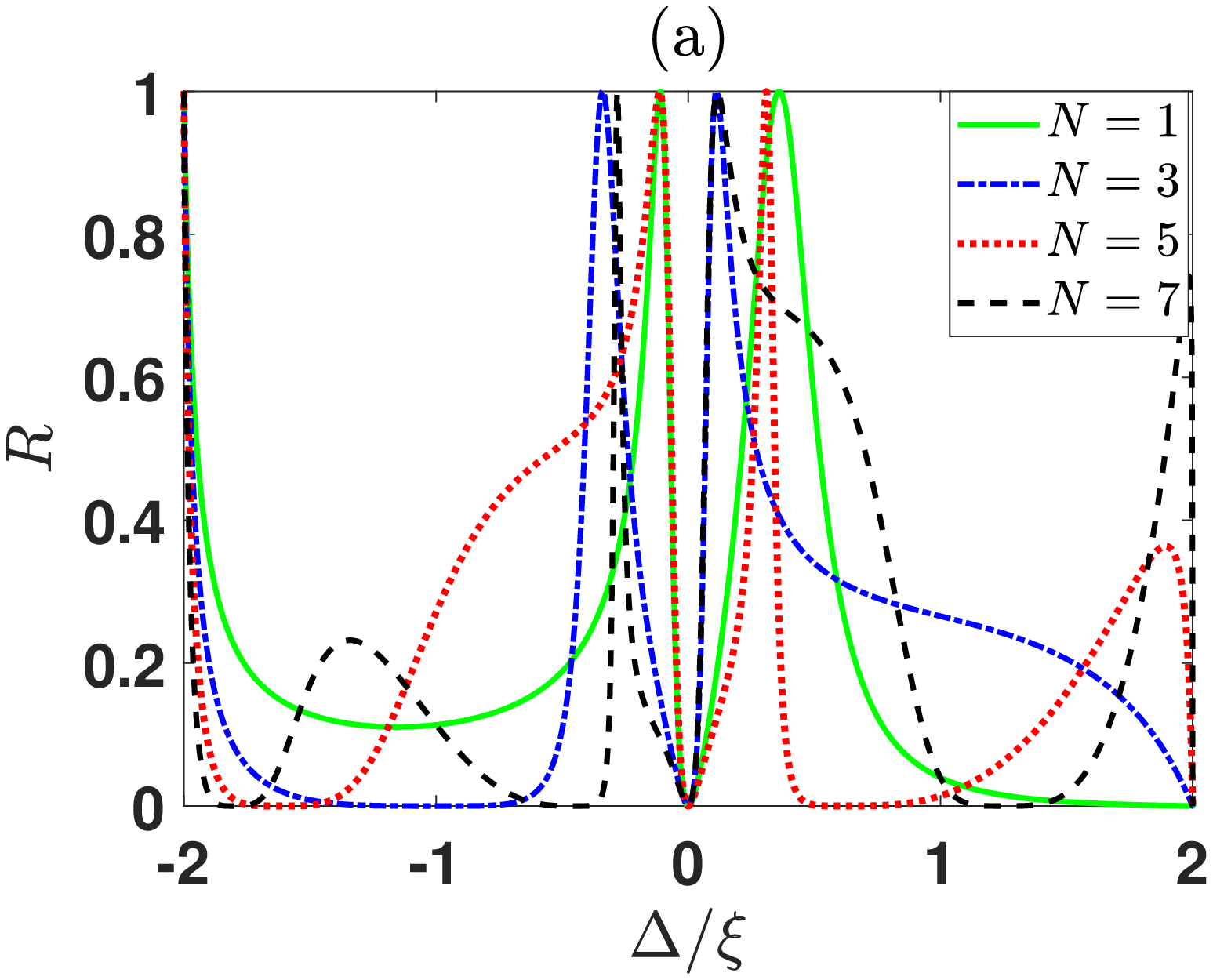}
\includegraphics[width=0.49\columnwidth]{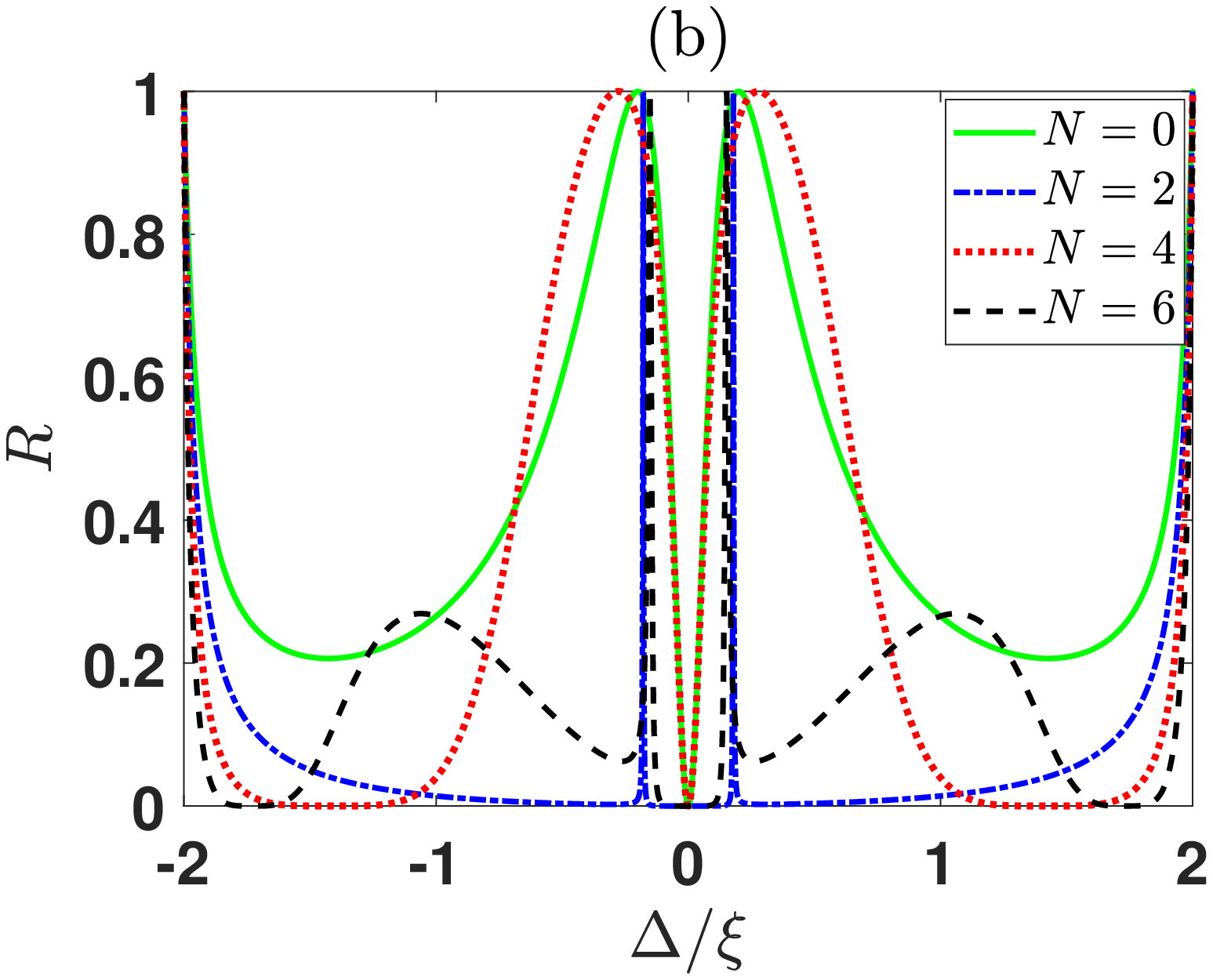}
\caption{The reflection rate $R$ as functions of detuning $\Delta$ for odd $N$ in (a) and even $N$ in (b). The parameters are set as $g=0.5\xi$, $\lambda=0.2\xi$,  $\omega _0=\omega _c=\Omega=20\xi$.}
\label{f2}
\end{figure}

In Fig.~\ref{f2}, we demonstrate the reflection rate $R$ as a function of the photon-atom detuning $\Delta$  by considering that the giant atom resonates with both the SAW resonaotor and the bare resonator in the CRW, that is, $\Omega=\omega_0=\omega_c$. Specifically, we illustrate the reflection rate when $N$ is odd and even in Fig.~\ref{f2} (a) and (b) respectively. For odd $N$, we find that the incident photon will completely be transmitted when it is resonant with the giant atom, that is $R=0$ when $\Delta=0$. It is dramatically different from the case without the SAW resonator, in which $R=0.5$ for $\Delta=0$~\cite{ZHAO2020}.  In this sense, the phonon in the SAW resonator can be used to modulate the photon scattering in the CRW.  In the regime for $\Delta\neq0$, it shows an asymmetry line type for the single photon reflection.

\begin{figure}
\centering
\includegraphics[width=0.49\columnwidth]{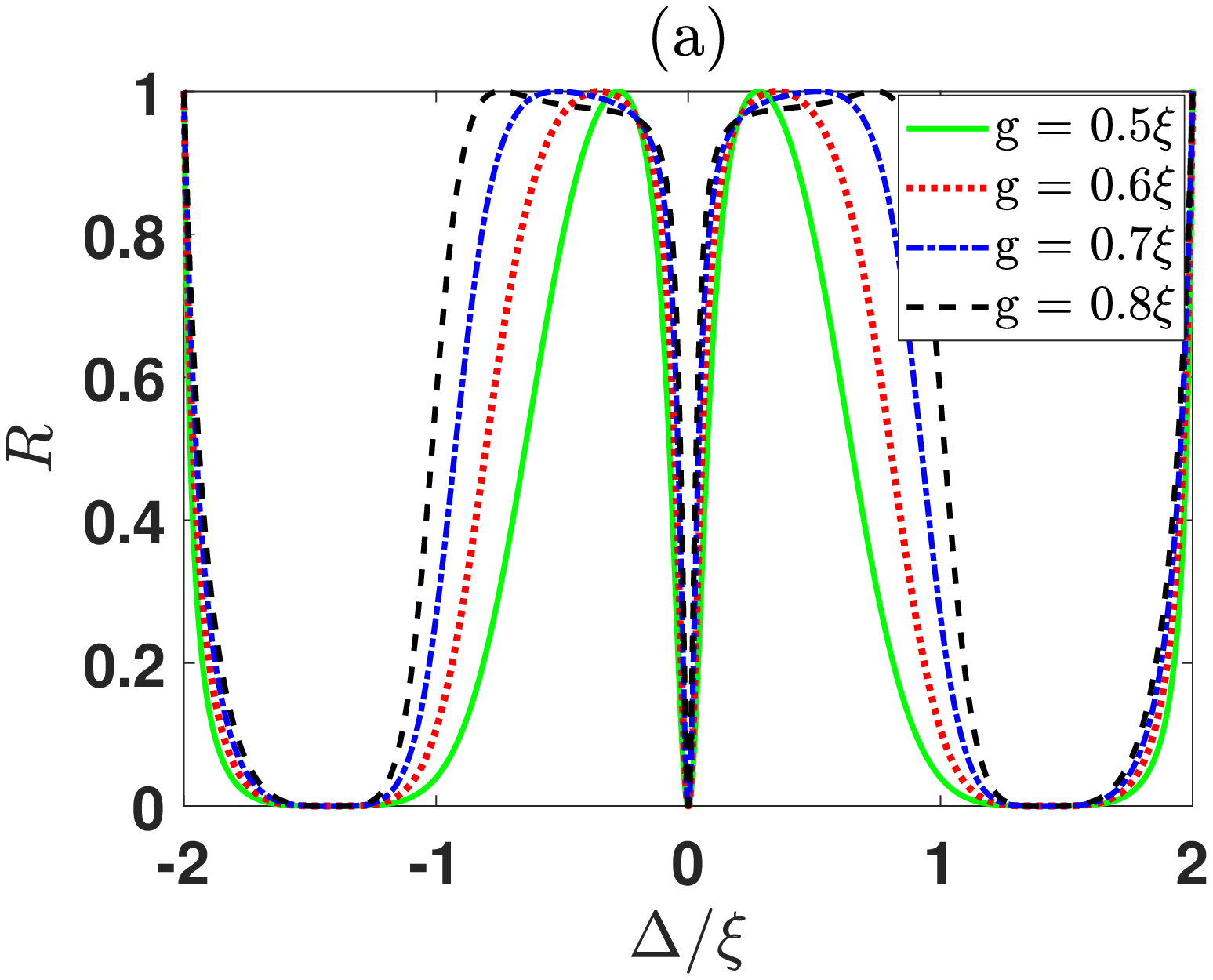}
\includegraphics[width=0.49\columnwidth]{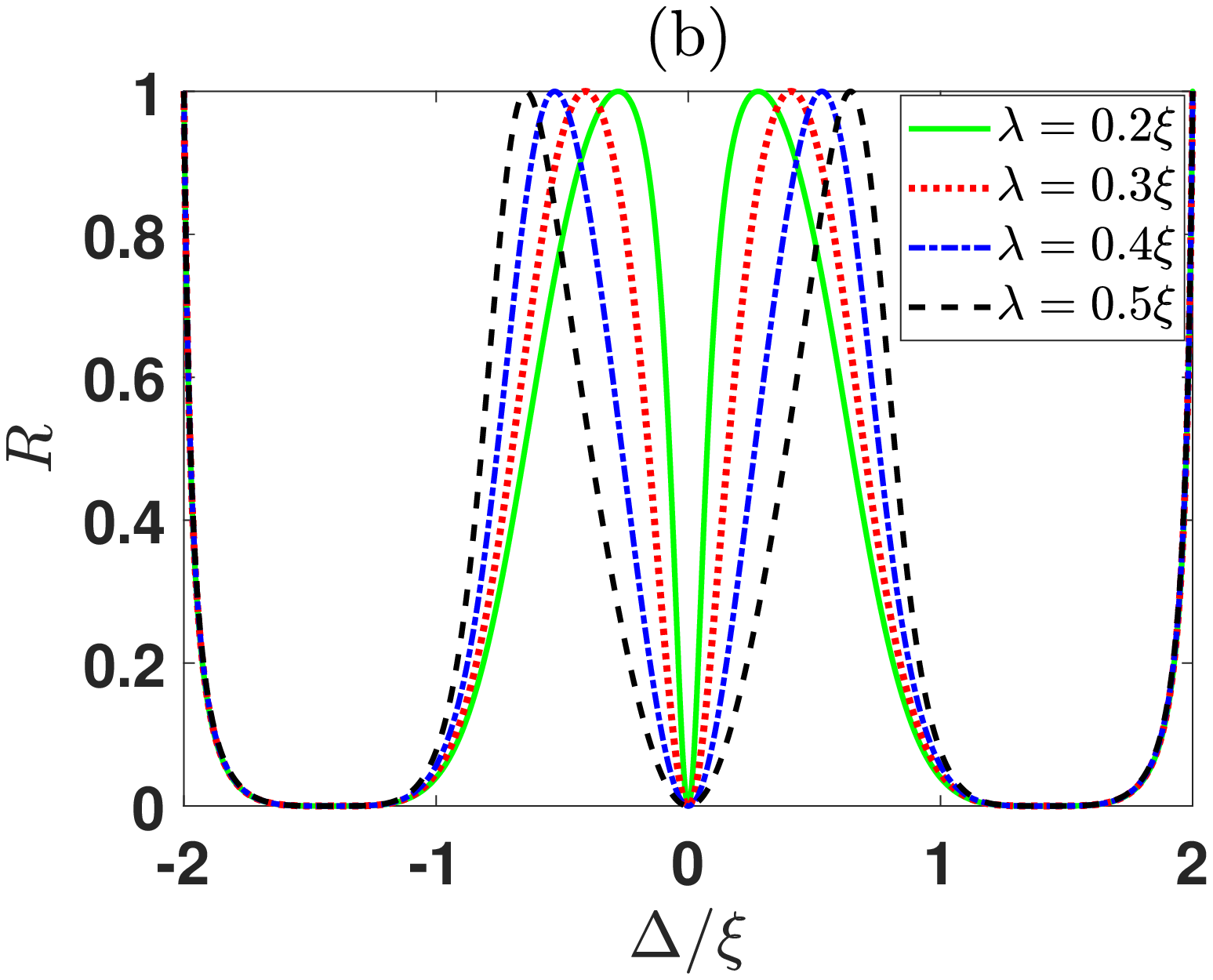}
\includegraphics[width=0.49\columnwidth]{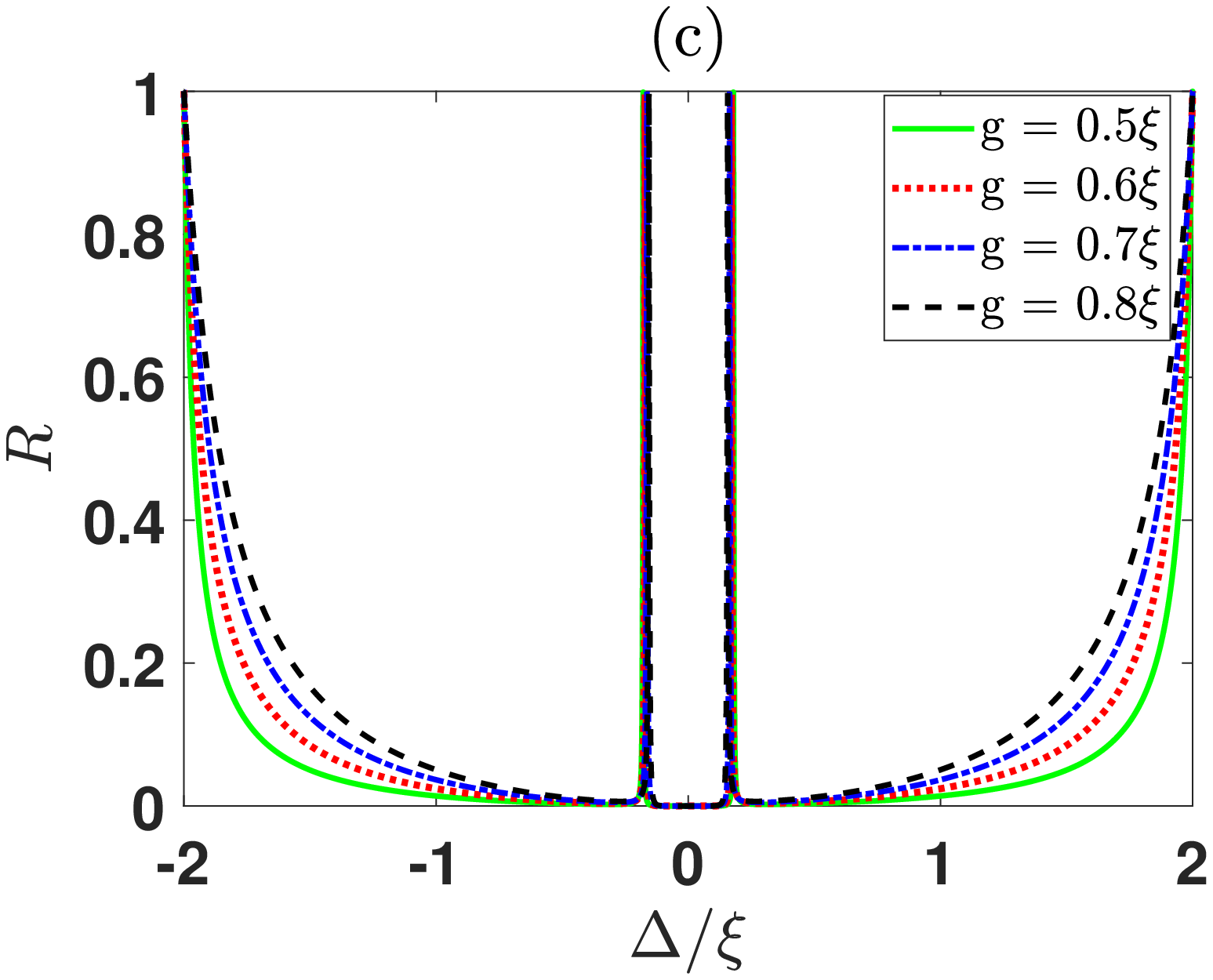}
\includegraphics[width=0.49\columnwidth]{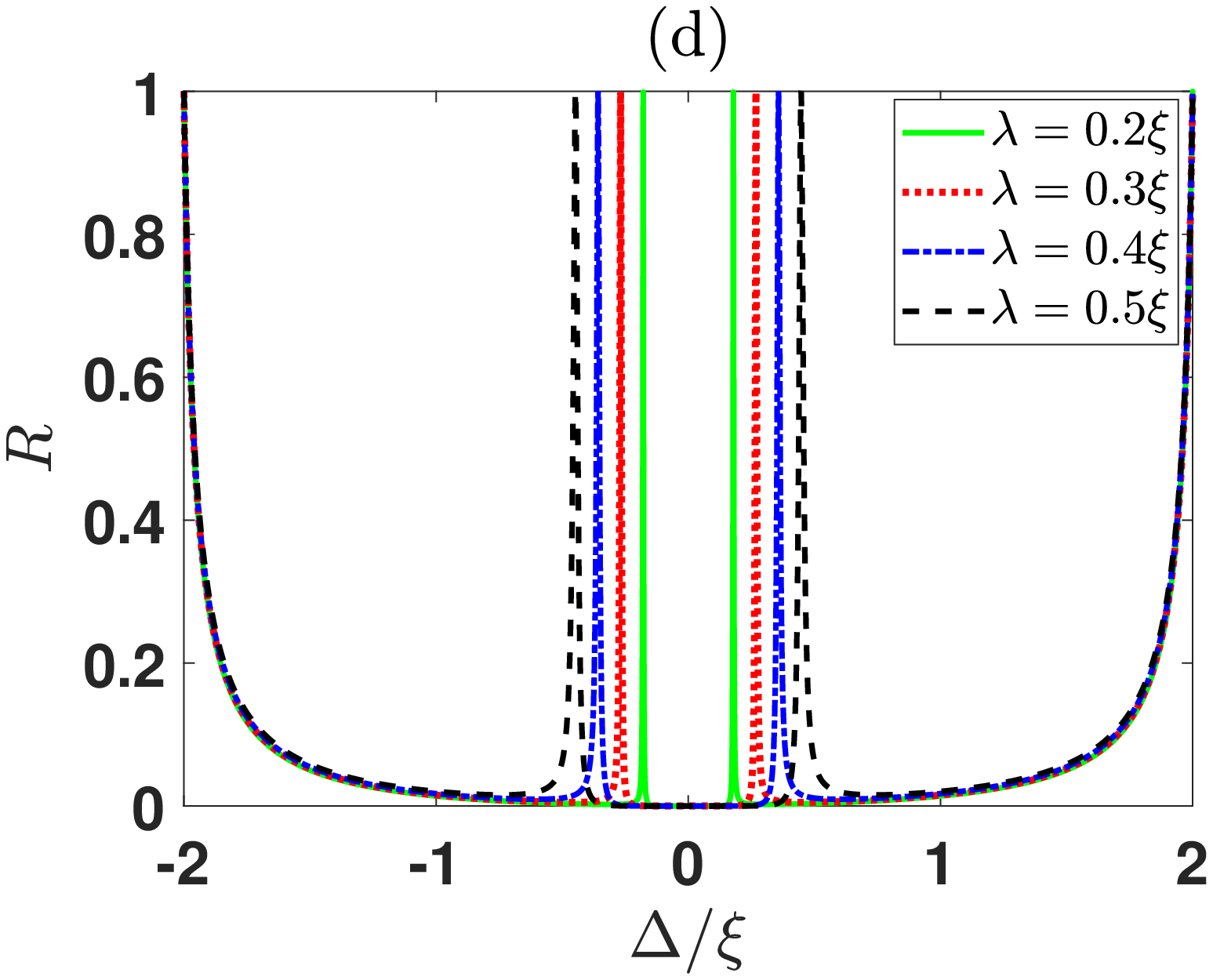}
\caption{The reflection rate $R$ for $N=4$ in panels (a,b) and  $N=2$ in panels (c,d). The parameters are set as  $\omega _0=\omega _c=\Omega=20\xi$, $\lambda=0.2\xi$ for panels (a) and (c), and $g=0.5\xi$ for panels (b) and (d) }
\label{f3}
\end{figure}

The modulation of the single photon scattering of the phonon in the SAW resonator is more interesting for the case of even $N$. As shown in Fig.~\ref{f2} (b), the peaks of the Rabi splitting experience a bit
shift in the giant atom setup compared to the small atom case with $N=0$. This shift is induced by the photon reflection via the two atom-CRW coupling sites. More interestingly, the narrow valley for $N=4m,\,m\in Z$ ($N=0,4,...$) is replaced by a relatively wide transmission  window for $N=4m+2$ ($N=2,6,...$). It is then necessary to investigate the effect of the atom-CRW and atom-SAW resonator coupling to the valley and window for the case of even $N$. We take $N=4$ and $N=2$ as the examples, we demonstrate the results in Fig.~\ref{f3}.  Comparing Fig.~\ref{f3} (a) with (b), we find that the atom-SAW resonator coupling strength $\lambda$ is more  effective at controlling the  width of the valley. That is, the width is nearly independent of the value of $g$ as shown in  Fig.~\ref{f3} (a), but a larger $\lambda$ will widen the valley obviously as shown in Fig.~\ref{f3} (b).  However, as shown in Fig.~\ref{f3} (a), the atom-CRW coupling can be used to widen the peaks which are induced by the Rabi splitting. This can be explained by considering the waveguide as a structured environment, and a stronger $g$ will induce a larger dissipation of the atom-SAW resonator dressed states, and is exhibited by the wider peaks. In Fig.~\ref{f3} (c) and (d), we also find that the width of the photonic transmission window is more sensitive to the atom-SAW resonator coupling strength $\lambda$ than of $g$.

\section{LARGER DETUNING}
\label{detuning}

\begin{figure}
\begin{centering}
\includegraphics[width=4cm]{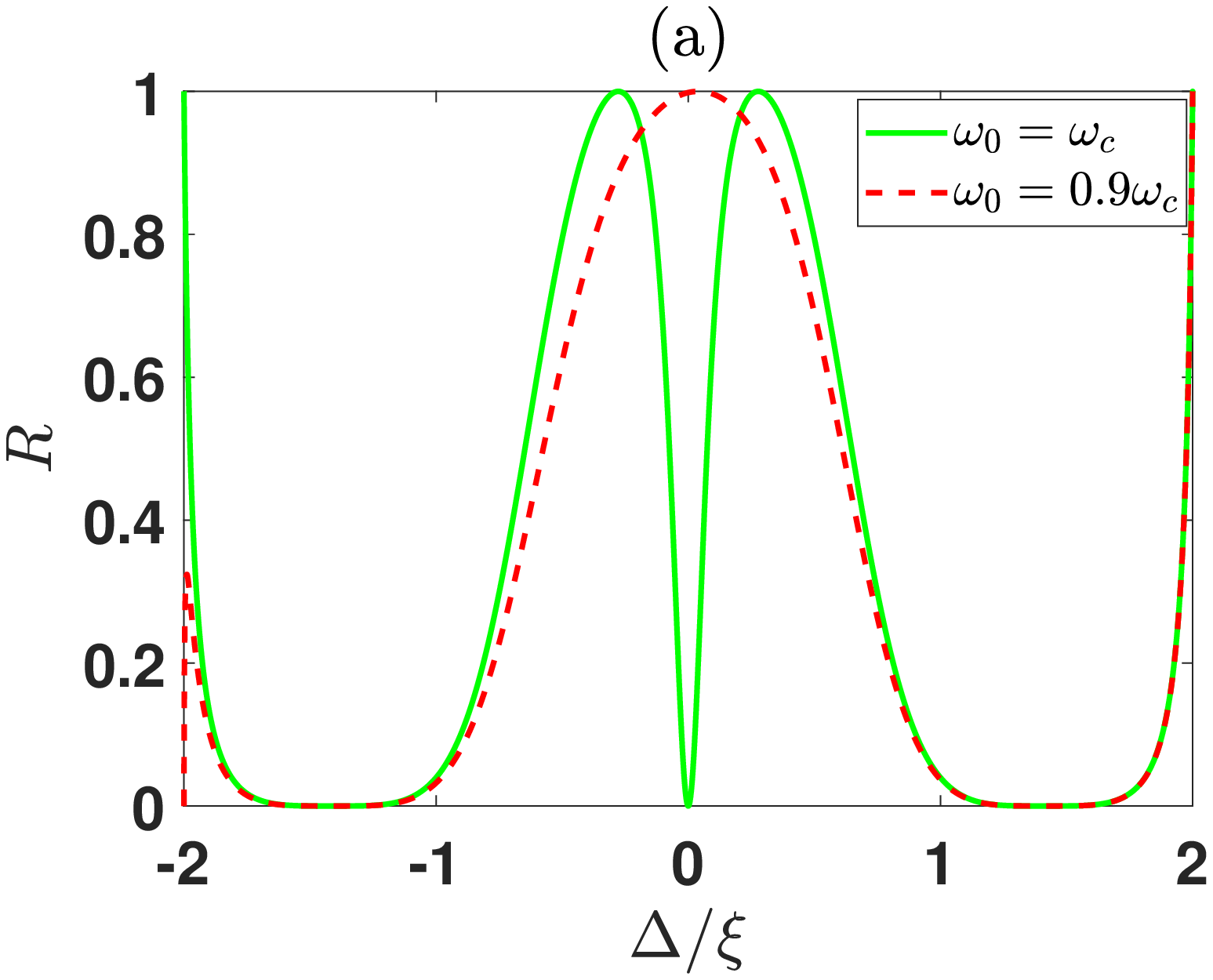}
\includegraphics[width=4cm]{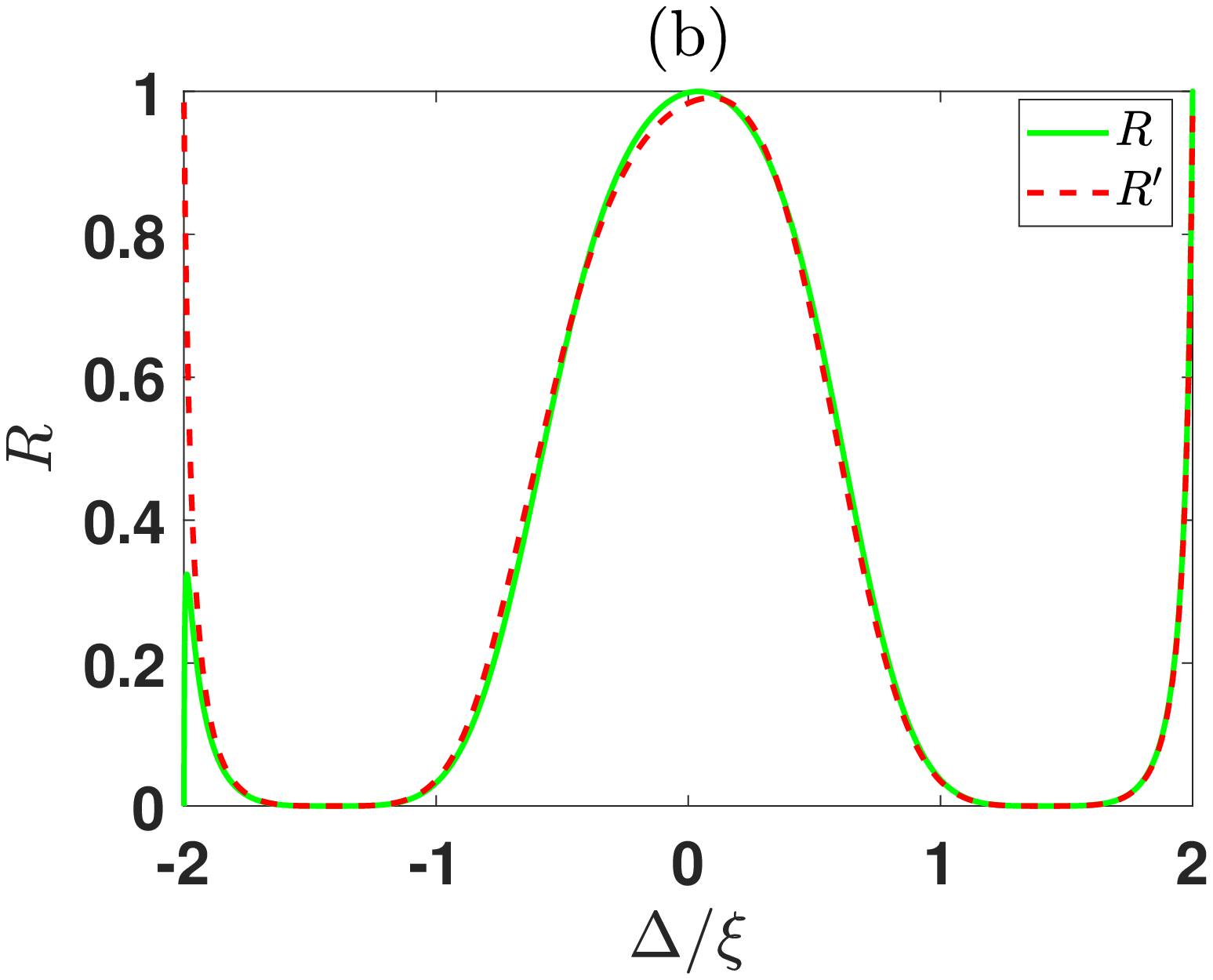}
 \par\end{centering}
 \caption{The single photon reflection spectrum. (a) The comparison of resonate and dispersive coupling. (b) The comparison between $R$ and $R^\prime$.
 The parameters are set as $N=4$, $g=0.5\xi$, $\lambda=0.2\xi$, $\omega _c=\Omega=20\xi$ and $\omega _0=0.9\omega _c$ in (b).}
  \label{f4}
\end{figure}

In the above discussions, we have found that the atom-SAW resonator coupling strength can be used to modulate the behavior of the single photon scattering in the waveguide when the atom is resonant with the SAW resonator. Since the detuning between the atom and the SAW resonator will change the nature of the effective coupling between them, it is expected the two peaks in the reflection spectrum will emerge into a single one as shown in Fig.~\ref{f4}(a). This can be explained by the dispersive coupling between the giant atom and the SAW resonator. In the case of large detuning $\lambda\ll|\omega_0-\Omega|$, we introduce the widely used Schrieffer-Wolff transformation~\cite{JRS1966,BRAVYI2011,HF1950,SN1953,LI2007} to derive the effective Hamiltonian $\mathcal{H}_{\rm eff}=e^{-S}\mathcal{H}e^{S}$, where the parent function is
\begin{equation}
\label{eq:9}
S=\frac{\lambda}{\Delta_c }(a\left | e \right \rangle\left \langle g \right |-a^\dagger\left | g \right \rangle\left \langle e \right | ),
\end{equation}
with $\Delta_c=\omega_{0}- \Omega$ being the detuning between the giant atom and the SAW resonator. Up to the second order of $\lambda/\Delta_c$, the effective Hamiltonian is obtained as
\begin{equation}
\label{eq:10}
\begin{split}
\mathcal{H}_{\rm eff}&=\omega_{0} a^{\dagger} a+\Omega|e\rangle\langle e|-\frac{\lambda^{2}}{\Delta_c}\left(a a^{\dagger}|e\rangle\langle e|-a^{\dagger} a| g\rangle\langle g|\right) \\
&+\omega_{c} \sum_{j} b_{j}^{\dagger} b_{j}-\xi \sum_{j=-\infty}^{+\infty}\left(b_{j+1}^{\dagger} b_{j}+b_{j}^{\dagger} b_{j+1}\right)\\
&+g\left[\left(b_{0}^{\dagger}+b_{N}^{\dagger}\right) \sigma^{-}+{\rm{H.c.}}\right]\\
&+\frac{\lambda^{2} g}{2 \Delta_c^{2}}\left[\left(b_{0}^{\dagger}+b_{N}^{\dagger}\right) \sigma^{-}+{\rm{H.c.}}\right]\\
&+\frac{\lambda}{\Delta_c} g\left[\left(b_{0}^{\dagger}+b_{N}^{\dagger}\right) a+{\rm{H.c.}}\right](|g\rangle\langle g|-| e\rangle\langle e|)\\
&+\frac{\lambda^{2} g}{\Delta_c^{2}}\left[\left(b_{0}^{\dagger}+b_{N}^{\dagger}\right) a+{\rm{H.c.}}\right] a \sigma^{+}\\
&+\frac{\lambda^{2} g}{\Delta_c^{2}}a^{\dagger}\left[\left(b_{0}^{\dagger}+b_{N}^{\dagger}\right) a+{\rm{H.c.}}\right] \sigma^{-}\
\end{split}
\end{equation}
Here, the first line represents the effective Hamiltonian of the atom-SAW resonator system, and the second line is the CRW Hamiltonian, which is not changed by the unitary transformation (same with $\mathcal{H}_c$), and the rest parts are the effective coupling between the atom-SAW resonator and the CRW.

Based on the effective Hamiltonian $\mathcal{H}_{\rm eff}$, we can still apply the analysis of the wave function in Eqs.~(\ref{eq:4},\ref{eq:5},\ref{eq:6}) to obtain the single photon reflection rate $R'$. However, it is too cumbersome to give the analytical expressions here. Therefore, we resort to a numerical calculation and illustrate the result in Fig.~\ref{f4}(b). As a comparison, we also plot the result of $R$. The good agreement between results shows the validity of the Schrieffer-Wolff transformation approach in the case of large detuning.

The Schrieffer-Wolff transformation supplies a way to understand why the two peaks are replaced by a single one in Fig.~\ref{f4} (a) in the large detuning. As shown by the first line of
Eq.~(\ref{eq:10}), which shows that the giant atom and the SAW resonator will not exchange excitation, that is, they form an effective dispersive coupling. As a result, we can not observe the Rabi splitting, which occurs in the resonantly coupling regime. One should note that the mechanism for the coalesce of the two peaks is completely different from that in Ref~\cite{ZENG2021}. In the later literature, the authors state that the coalesce is induced by the fact that the high temperature destroys the quantum nature of the system.

\section{REMARKS and CONCLUSIONS}

\label{conclusion}

The giant atom setup has been experimentally realized recently~\cite{MVGUS2014,Kannan2020,AMV2021,WANGZQ2022}, where the giant atom is served by superconducting qubit or magnon spin ensemble.  With the available technologies, the coupling between the SAW and the superconducting qubit has been realized with coupling strength $\lambda/(2\pi)\approx 20$~MHz~\cite{ZENG2021}, the controllable CRW has also been realized by the high-impedance microwave resonators and the nearest hopping strength has been achieved by $\xi\approx 200$~MHz. The qubit-resonator coupling strength is approximately $300$~MHz~\cite{PRX2022} and a weaker coupling $\lambda=0.5\xi$ is undoubtedly available.  The loss of the qubit as well as the phonon and photon mode are in the regime of tens of kHz~\cite{ZENG2021,PRX2022}, which is three orders weaker than the above coupling strength, and is therefore neglected here.

In this paper, we have demonstrated how to control the scattering of the photon by utilizing the phonon with the assistance of the giant atom setup. The giant atom has been realized experimentally in superconducting circuits, where the superconducting qubit or spin ensemble in magnon couples to the transmission line via more than one connecting point. On the other hand, SAW resonator has been widely used to design  microelectro mechanical devices~\cite{GUO2017,AFK2014,MDS2007}, and the current technology has brought into the quantum regime, which invokes the topics of circuit quantum acoustic dynamics (cQAD). The superconducting qubits have provided a medium to realize the transform between the phonons and photons, within or outside the same frequency regime. We hope that our work on controlling photons via phonon will stimulate new studies of the hybrid system and  broaden the application of artificial giant atom.

\begin{acknowledgments}
This is supported by National Key R\&D Program of China (Grant No. 2021YFE0193500), Science and Technology Development Project of Jilin Province (Grant No. 20230101357JC)
\end{acknowledgments}


\end{document}